\documentclass[preprint,12pt]{elsarticle}

\usepackage{graphicx}
\usepackage{enumitem}
\usepackage{amssymb}

\usepackage{geometry}
\usepackage{caption}
\usepackage{makecell}
\usepackage{tabularx}
\usepackage[title]{appendix}

\newcounter{bla}

\journal{Computer Physics Communications}

\begin{document}

\begin{frontmatter}



\title{\texttt{phq}: a Fortran code to compute phonon quasiparticle properties and dispersions}


\author[a]{Z. Zhang}
\author[b,c]{D.-B. Zhang\corref{author}}
\author[d] {T. Sun}
\author[a,e]{R. M. Wentzcovitch\corref{author}}

\cortext[author] {Corresponding author.\\\textit{E-mail address:} dbzhang@csrc.ac.cn; tsun@ucas.ac.cn; rmw2150@columbia.edu }
\address[a]{Department of Applied Physics and Applied Mathematics, Columbia University, New York, New York 10027, USA}
\address[b]{Beijing Computational Science Research Center, Beijing 100193, China}
\address[c]{College of Nuclear Science and Technology, Beijing Normal University, Beijing 100875, China}
\address[d]{Key Laboratory of Computational Geodynamics, Chinese Academy of Sciences, Beijing 100049, China}
\address[e]{Department of Earth and Environmental Sciences, Lamont-Doherty Earth Observatory, Columbia University, New York, New York 10964, USA}

\begin{abstract}
We here introduce a Fortran code that computes anharmonic free energy of solids from first-principles based on our phonon quasiparticle approach. In this code, phonon quasiparticle properties, i.e., renormalized phonon frequencies and lifetimes, are extracted from mode-projected velocity auto-correlation functions (VAF) of modes sampled by molecular dynamics (MD) trajectories. Using renormalized frequencies as input, the code next constructs an effective harmonic force constant matrix to calculate anharmonic phonon dispersions over the whole Brillouin zone and thus the anharmonic free energy in the thermodynamic limit ($N \rightarrow \infty$). We illustrate the use of this code to compute \textit{ab initio} temperature-dependent anharmonic phonons of Si in the diamond structure.
\end{abstract}


\end{frontmatter}


\noindent
{\bf PROGRAM SUMMARY}
\\
\begin{small}
{\em Program title:} \texttt{phq}                                         \\
{\em Licensing provisions:} GPLv3                                   \\
{\em Programming language:} Fortran 90                                  \\
{\em Operating system:} Linux, macOS, Windows                               \\
{\em Nature of problem:}\\
Accurate free energy calculations deliver predictive thermodynamic properties of solids. Although the quasi-harmonic approximation (QHA) has been widely used for various materials, its validity at very high temperatures is not guaranteed because of intrinsic anharmonic effects. The QHA cannot be used even at low temperatures for strongly anharmonic systems, e.g., those that are stabilized by anharmonic fluctuations. When anharmonicity is non-negligible, phonon frequencies display a pronounced dependence on temperature, which impacts thermodynamic properties. Therefore, the calculation of anharmonic phonon dispersions is critical for an accurate estimation of the free energy. \\
{\em Solution method:}\\
We calculate phonon quasiparticle properties, i.e., renormalized phonon frequencies and lifetimes, by combing molecular dynamics (MD) trajectories and harmonic normal modes (HNM). MD velocities are projected into HNMs and the mode-projected velocity auto-correlation functions (VAF) are then calculated.  Quasiparticle properties are then obtained by analyzing the VAFs, if they are well defined. Using the renormalized phonon frequencies and HNMs we build an effective (temperature-dependent) harmonic force constant matrix from which the anharmonic phonon dispersions can be obtained. \\
{\em Unusual features:}\\
(i) Because MD captures anharmonic effects to infinite order, this approach to compute renormalized phonon frequencies and lifetimes is expected to be more appropriate than lowest-order perturbation theory for strongly anharmonic systems or for weakly anharmonic systems at very high temperatures.
(ii) By characterizing phonon quasiparticles, this approach provides a validity check of the phonon concept, a fundamental procedure missing in other approaches to compute anharmonic dispersions, e.g., the self-consistent phonon method.
\\


\end{small}


\section{Introduction}

The phonon gas model (PGM)~\cite{pgm1, Sun2010, pgm2, pgm3} provides an excellent paradigm for calculating thermodynamics properties of materials. For weakly anharmonic systems, anharmonic effects reflecting the intrinsic temperature dependence of phonon frequencies are negligible. Thus, the quasiharmonic approximation (QHA) \cite{qha1, qha2} offers predictive free energies and thermodynamics properties. Note that the QHA accounts for the extrinsic temperature dependence of the phonon frequencies caused by volume changes (thermal expansion). However, in strongly anharmonic systems, intrinsic anharmonic effects are key to structural stabilization and must be properly accounted for. Therefore, anharmonic phonon dispersions are required for estimations of free energy and related thermodynamic properties.

Here we introduce a code of a hybrid approach which fully accounts for anharmonic effects. In this approach, we describe a phonon quasiparticle of normal mode $({\bf q},s)$ by the mode-projected velocity autocorrelation functions (VAF) \cite{vaf},
\begin{equation}
\langle V_{{\bf q}s}(0)\cdot V_{{\bf q}s}(t) \rangle = \lim_{\tau \rightarrow \infty} \frac{1}{\tau} \int_{0}^{\tau} V^*_{{\bf q}s}(t') V_{{\bf q}s}(t'+t) dt'.
\end{equation}
where,
\begin{equation}
V_{{\bf q}s}(t) = \sum _{i=1} ^N V(t) \cdot e^{i {\bf q} \cdot {\bf r}_i} \cdot {\bf \hat{e}}_{{\bf q}s}
\end{equation}
is the $s$-th branch of {\bf q}-projected velocity. ${\bf \hat{e}}_{{\bf q}s}$ is the harmonic phonon polarization vector of mode (${\bf q},s$).
\begin{equation}
V(t) = V(\sqrt {M_1} {\bf v}_1(t), ..., \sqrt {M_N} {\bf v}_{N}(t))
\end{equation}
is the weighted velocity with $3N$ components. ${\bf v}_i(t)(i=1,...,N)$ are atomic velocities calculated from MD trajectories of an $N$-atom supercell. $M_i$ is the atomic mass of the $i^{th}$ atom in the supercell.

For a well-defined phonon quasiparticle, its power spectrum,
\begin{equation}
G_{{\bf q}s}(\omega) = \left|\int_{0}^{\infty} \langle V_{{\bf q}s}(0)\cdot V_{{\bf q}s}(t) \rangle e^{i \omega t} dt\right|^2.
\end{equation}
should have a Lorentzian line shape with a peak at $\widetilde{\omega}_{{\bf q}s}$ and a linewidth of 1/(2$\tau_{{\bf q}s}$), where $\tau_{{\bf q}s}$ is the lifetime of mode (${\bf q},s$) in the relaxation time approximation \cite{Sun2010, pgm2,  HofScience}. In principle, a complete decay of $\langle V_{{\bf q}s}(0)\cdot V_{{\bf q}s}(t) \rangle$ is required to obtain a reliable $G_{{\bf q}s}$. However, very long MD runs are needed to satisfy this condition for phonons with long lifetimes. This is inconvenient for \textit{ab initio} MD simulations. Here, we extract  $\widetilde{\omega}_{{\bf q}s}$ and $\tau_{{\bf q}s}$ from $\langle V_{{\bf q}s}(0)\cdot V_{{\bf q}s}(t) \rangle$ phenomenologically from relatively shorter MD runs. For a well-defined phonon quasiparticle of mode (${\bf q},s$), one can simply fit $\langle V_{{\bf q}s}(0)\cdot V_{{\bf q}s}(t) \rangle$ to the expression,
\begin{equation}
A_{{\bf q}s}{\rm cos}(\widetilde{\omega}_{{\bf q}s} t) e^{-t/(2\tau_{{\bf q}s})}
\end{equation}
where, $A_{{\bf q}s}$ is the vibrational amplitude. In practice, to obtain reliable $\widetilde{\omega}_{{\bf q}s}$ and $\tau_{{\bf q}s}$, the fitting needs only the numerical data of $\langle V_{{\bf q}s}(0)\cdot V_{{\bf q}s}(t) \rangle$ for the first few oscillation periods.

With $\widetilde{\omega}_{{\bf q}s}$ of those modes sampled by the MD run, an effective harmonic dynamical matrix can be constructed,
\begin{equation}
\widetilde{D}({\bf q}) = \hat{{\bf e}}_{\bf q} \Omega_{\bf q} \hat{{\bf e}}_{\bf q}^\dagger,
\end{equation}
where,
\begin{equation}
\Omega_{\bf q} = {\rm diag}[\widetilde{\omega}^2_{{\bf q}1}, \widetilde{\omega}^2_{{\bf q}2}, ..., \widetilde{\omega}^2_{{\bf q}3N}].
\end{equation}
and $\hat{{\bf e}}_{\bf q}$ is the harmonic eigenvector matrix:
\begin{equation}
\hat{{\bf e}}_{\bf q} = [\hat{{\bf e}}_{{\bf q}1}, \hat{{\bf e}}_{{\bf q}2}, ..., \hat{{\bf e}}_{{\bf q}3N}].
\end{equation}
The effective harmonic force constant matrix $\widetilde{\Phi}({\bf r})$ can be obtained from the Fourier transformation of $\widetilde{D}({\bf q})$, from which,
\begin{equation}
\widetilde{D}({\bf q'}) = \sum_{\bf r} \widetilde{\Phi}({\bf r}) \cdot e^{-i{\bf q'} \cdot {\bf r}}
\end{equation}
the effective harmonic dynamical matrix at arbitrary wave vector ${\bf q'}$ is obtained. This way, $\widetilde{\omega}_{{\bf q'}s}$ at any ${\bf q'}$ in the Brillouin zone are calculated by diagonalizing $\widetilde{D}({\bf q'})$.

This approach gives anharmonic phonon dispersions that are useful to study thermodynamic and lattice thermal transport properties of materials, especially at extreme conditions. It also offers a means of checking the existence of feasible vibrational modes even in the presence of strong anharmonicity. For example, we can witness the emergence of the cubic phase of CaSiO$_3$ perovskite at high temperature and high pressure \cite{Sun2014}. The concept of phonon quasiparticle also helps to easily resolve mode frequencies in the vibrational spectrum that overlap in energy even in the presence of crystal structure complexity. This advantage has been demonstrated on MgSiO$_3$ perovskite, a typical non-trivial crystal structure with \textit{Pbnm} symmetry and 20 atoms/cell. The obtained subtle and irregular frequency shifts with temperature are in good agreement with experiments \cite{Zhang2014}. Recently, we applied this approach to beryllium metal to address phase transitions at temperatures near melting. Our calculations confirmed that the long-debated and controversial \textit{hcp}$\rightarrow$\textit{bcc} transition occurs at very high temperatures, just before melting \cite{Lu2017}. While the \textit{hcp} structure was shown to be dynamically stable at all temperatures up to the transition temperature, the \textit{bcc} phase was dynamically stabilized by anharmonic fluctuations just below the transition temperatures. At the phase boundary, both structures had phonon quasiparticles well defined and that was sufficient to enable free energy and phase boundary computations at those high temperatures. In addition, phonon lifetimes obtained this way allow us to study lattice thermal conductivity using the Boltzmann transport equation (BTE) approach. For instance, we demonstrated for the first time the breakdown of the well-known phonon minimal mean free theory in MgSiO$_3$ perovskite~\cite{Zhang2017}, a result later confirmed by lowest-order perturbation theory~\cite{Ghaderi2017}.

The procedure to compute quasiparticle properties consists of four key steps. Step 1, atomic velocities are calculated by carrying out MD simulations and recording velocities [Eq.~(3)]. Step 2, projecting instantaneous atomic velocities into the harmonic normal modes (HNM) [Eq.~(2)]. Step 3, computing mode-projected velocity auto-correlation functions (VAF) and extracting quasiparticle frequencies and lifetimes from VAFs [Eq.~(1)(4)(5)]. Step 4, constructing effective harmonic dynamical matrices using the renormalized phonon frequencies and HNMs and using them to obtain anharmonic phonon dispersions [Eq.~(6)(7)(8)(9)].

In Step 3, three numerical techniques are provided to obtain quasiparticle properties: (i) Fitting the VAFs to an analytical function, Eq.~(5), (ii) Fourier transforming (FT) Appendix \ref{appendix:ft}~\cite{recipes} VAFs, and (iii) using the maximum entropy method (MEM) Appendix \ref{appendix:mem}~\cite{recipes}. Also, we note that Step 4, critical to obtaining anharmonic phonon dispersions and free energies, is missing from a similar work \cite{dynaphopy}.

\section{Description of the program}

The \texttt{phq} package folder is contained in the compressed \texttt{phq.zip} file. Inside the folder, there are \texttt{LICENSE} file, \texttt{README.md} file and three sub-folders. There is an example of diamond Si in the \texttt{example} sub-folder. Four makefiles are in the \texttt{system} sub-folder. \texttt{makefile\_linux} is for Linux or macOS operating systems and \texttt{makefile\_windows} is for Windows operating system. Makefiles for both \texttt{gfortran} and \texttt{ifort} compilers are also prepared. \texttt{ifort} compiler is recommended for faster running speed and smaller memory requirement. One can modify properly the \texttt{FC} flag to enable other compilers.

The source code is in the \texttt{src} sub-folder with three Fortran90 files: \texttt{configure.f90} contains three auxiliary modules, which configure the physical quantities, parameters and text read in settings. \texttt{main.f90} is the main module where: (i) the mode-projected velocities are calculated from the MD trajectory; (ii) renormalized phonon frequencies and lifetimes are extracted from the mode-projected VAFs; (iii) effective harmonic force constants and thus anharmonic phonon dispersions are obtained.  \texttt{phq.f90} runs the program and records the running time. A detailed description of subroutines used in the \texttt{main.f90} file is listed in Table \ref{table:main}.

\begin{table}[]
  \caption{A description of subroutines in \texttt{main.f90}}
  \centering
  \renewcommand{\arraystretch}{1.2}
  \begin{tabularx}{\textwidth}{l X}
    \Xhline{4\arrayrulewidth}
    Subroutine name & Description  \\
    \Xhline{3\arrayrulewidth}
    \texttt{control} & General control of the program. \\
    \hline
    \texttt{read\_scf} & Read in structural information of the primitive cell. \\
    \hline
    \texttt{read\_md} & Read in MD information and compute atomic displacements from their static positions. \\
    \hline
    \texttt{properties} & Compute the heat capacity via heat fluctuation. If no information other than atomic coordinates is provided in the \texttt{md.out} file, comment out this subroutine in the control subroutine. \\
    \hline
    \texttt{read\_dyn} & Read in harmonic phonon information, compute eigenmodes of the supercell and the primitive cell at different {\bf q}-point and check the orthogonality. \\
    \hline
    \texttt{harmonic\_matrix} & Compute the harmonic force constant matrix from harmonic frequencies and harmonic eigenmodes. \\
    \hline
    \texttt{harmonic force} & Compute the interatomic harmonic forces. \\
    \hline
    \texttt{frequency} & Compute the atomic velocities and perform velocity eigenmode projection, according to Eq.~(2)(3). \\
    \hline
    \texttt{equipartition} & Compute the ionic kinetic energy from equipartition theorem. \\
    \hline
    \texttt{correlation} & Compute the VAFs, accourding to Eq.~(1). \\
    \hline
    \texttt{correlation\_fit} & Fit the VAFs to Eq.~(5) and obtain quasiparticle properties~\cite{Sun2010}. \\
    \hline
    \texttt{corr\_fourier} & Perform FT of VAFs according to Eq.~(4) and obtain quasiparticle properties. \\
    \hline
    \texttt{corr\_fit\_fourier} & Perform FT of the fitted curves. Avoid calling this subroutine if shorter running time is desired. \\
    \hline
    \texttt{maximum\_entropy} & Perform MEM calculation and obtain the quasiparticle properties. \\
    \hline
    \texttt{linear\_response} & Obtain MEM power spectrum via linear prediction method. \\
    \hline
    \texttt{lorentzian} & Obtain quasiparticle properties by performing a fitting of the MEM power spectrum to a Lorentzian function.\\
    \hline
    \texttt{gamma\_matrix} & Compute the effective harmonic force constant matrix from renormalized frequencies and harmonic eigenmodes. \\
    \hline
    \texttt{gamma\_force} & Compute the interatomic anharmonic forces. \\
    \hline
    \texttt{dynamics\_matrix} & Compute the harmonic dynamical matrices from harmonic frequencies and harmonic eigenmodes. Comment out this subroutine if the effective harmonic dynamical matrices are desired. \\
    \hline
    \texttt{dynamics\_matrix\_md} & Compute the effective harmonic dynamical matrices from renormalized frequencies and harmonic eigenmodes, according to Eq.~(6)(7)(8). \\
    \hline
    \texttt{write\_dym} & Write down the effective harmonic dynamical matrices or the harmonic dynamical matrices with the same format of \texttt{ph.x} executable's output, which can be read in by \texttt{q2r.x} executable in the Quantum ESPRESSO~\cite{qe} suite to perform Eq.~(9). \\
    \Xhline{4\arrayrulewidth}
  \end{tabularx}
  \label{table:main}
\end{table}

\begin{figure}[h]
  \centering
  \includegraphics[width=1.0\linewidth]{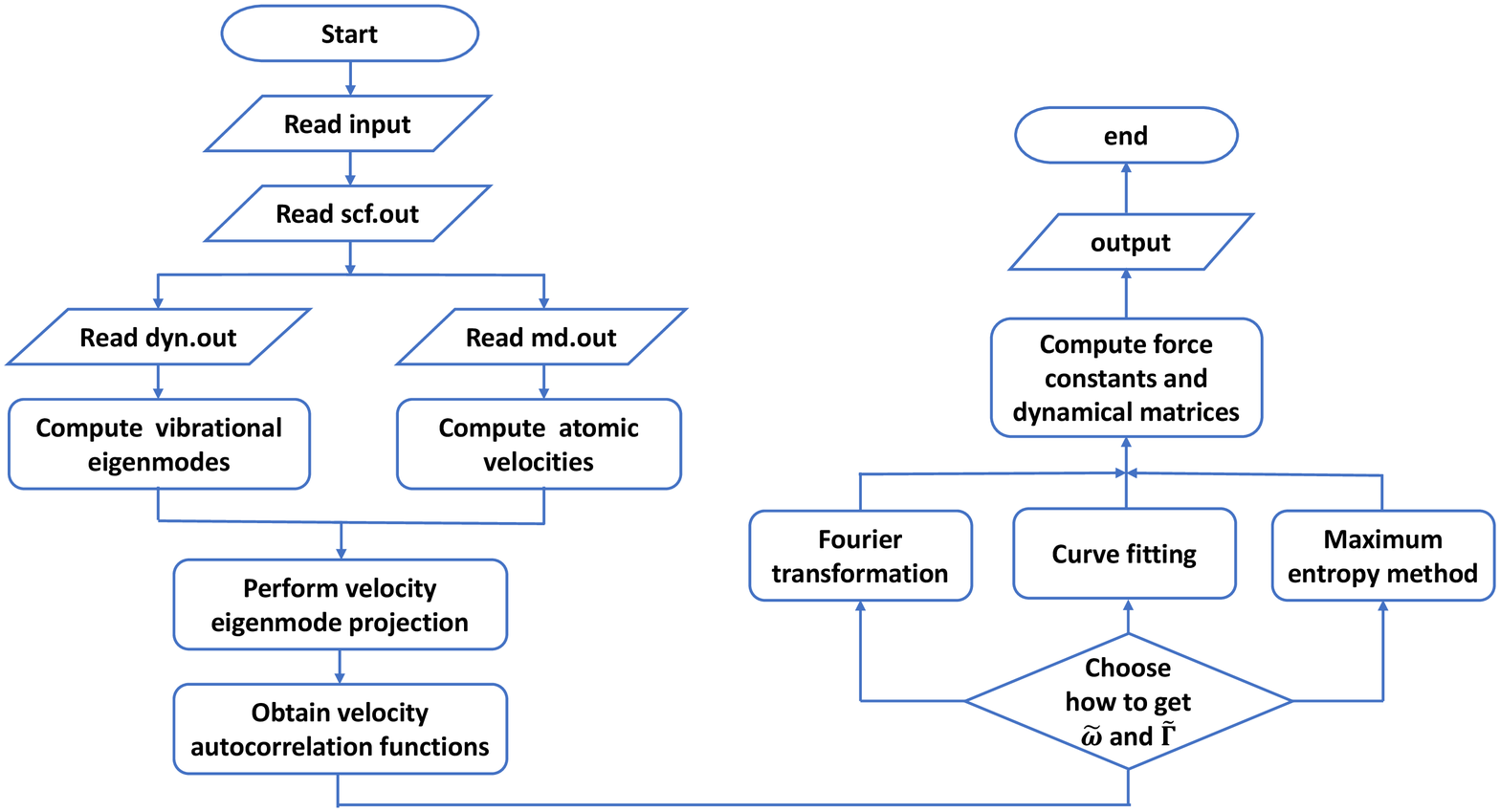}
  \caption{Flow chart of the \texttt{phq} program}
  \label{fig:workflow}
\end{figure}

After making the file, users need to put the \texttt{phq} executable into the same folder of those four input files in order to run the code. Run command \texttt{./phq < input} to perform the calculation. The input files will be introduced in the {\bf Input files} section. The code workflow is shown in Fig. \ref{fig:workflow}.

\section{Input files}
There are 4 input files for the \texttt{phq} program: \texttt{input}, \texttt{scf.out}, \texttt{dyn.out} and \texttt{md.out}, which provide the general settings, structure, harmonic phonon, and MD information separately. Users can prepare the input files according to the \texttt{example} provided within the code package. The instructions for preparing the input files are as following:

\subsection{input}
\texttt{input} provides general settings for the calculation. The values of the parameters should be set by the user. The details of the parameters are described in Table \ref{table:input}.

\begin{table}[]
  \caption{Parameters used in the \texttt{input}}
  \centering
  \renewcommand{\arraystretch}{1.2}
  \begin{tabularx}{\textwidth}{l X}
    \Xhline{4\arrayrulewidth}
    Parameter name & Description \\
    \Xhline{3\arrayrulewidth}
    \texttt{dt} & MD time step in atomic unit. If 1 fs time step is used, simply enter dt = 20.67055273. \\
    \hline
    \texttt{step\_md\_use} & Number of MD steps needed for the calculation of mode-projected VAFs. \\
    \hline
    \texttt{correlation\_time} & Desired decay time for VAFs in unit of \texttt{dt}. \\
    \hline
    \texttt{pole} & Parameter used in the MEM. It is used to filter high-frequency components. Usually a reasonable range of pole (200 $\sim$ 2000) can yield smooth spectrum. \\
    \hline
    \texttt{supercell} & Supercell used in MD related to the primitive cell. 3 integers are required to enter. \\
    \hline
    \texttt{temperature} & MD temperature in unit of Kelvin. \\
    \hline
    \texttt{method} & Enter one of the integer number 0, 1 or 2 to select renormalized frequencies obtained by which method to compute the effective harmonic force constants and dynamical matrices. 0 represents the fitting approach (Eq.~(5)). 1 represents the Fourier transformation. 2 represents the maximum entropy method. \\
    \Xhline{4\arrayrulewidth}
  \end{tabularx}
  \label{table:input}
\end{table}

\subsection{scf.out}
\texttt{scf.out} provides structure information of the primitive cell from self-consistent calculations. The values of the parameters should be same as those used in first-principles calculations. The details of the parameters are described in Table \ref{table:scf}.

\begin{table}[]
  \caption{Parameters used in the \texttt{scf.out}}
  \centering
  \renewcommand{\arraystretch}{1.2}
  \begin{tabularx}{\textwidth}{l X}
    \Xhline{4\arrayrulewidth}
    Parameter name & Description \\
    \Xhline{3\arrayrulewidth}
    \texttt{ntype} & Number of different elements in the primitive cell. \\
    \hline
    \texttt{natom} & Number of atoms in the primitive cell. \\
    \hline
    \texttt{mass} & Atomic mass of each element following the symbol of the element. \\
    \hline
    \texttt{lattice\_parameter} & Scale of lattice vectors in unit of Bohr radius. \\
    \hline
    \texttt{cell\_parameters} & Lattice vectors in cartesian cooradinate in unit of \texttt{lattice\_parameter}. \\
    \hline
    \texttt{atomic\_positions} & Atomic positions of atoms in the primitive cell in reduced coordinates. \\
    \Xhline{4\arrayrulewidth}
  \end{tabularx}
  \label{table:scf}
\end{table}

\subsection{dyn.out}
\texttt{dyn.out} provides harmonic phonon information. The format follows the \texttt{ph.x} executable's output of the Quantum ESPRESSO \cite{qe} suite. Quantum ESPRESSO users can simply obtain the \texttt{dyn.out} file by: (i) Turn off the symmetry in the self-consistent calculations by \texttt{pw.x}. (ii) Do the harmonic phonon calculations by \texttt{ph.x}. (iii) Concatenate \texttt{dynmat1}, ..., \texttt{dynmatnq} together, where nq is the total number of the {\bf q}-point. The details of the parameters are described in Table \ref{table:dyn}.

\begin{table}[]
  \caption{Parameters used in the \texttt{dyn.out}}
  \centering
  \renewcommand{\arraystretch}{1.2}
  \begin{tabularx}{\textwidth}{l X}
    \Xhline{4\arrayrulewidth}
    Parameter name & Description \\
    \Xhline{3\arrayrulewidth}
    \texttt{q} & {\bf q}-point in cartesian coordinates in unit of 2$\pi$/\texttt{lattice\_parameter}. This parameter is required.\\
    \hline
    \texttt{freq} & Harmonic phonon frequencies. This parameter in unit of THz is required as program input as well as the following six columns of eigenvectors. Eigenvector of each atom has x, y and z components and each component has a real part and an imaginary part. The order of atoms should be the same as that of the atoms entered in \texttt{atomic\_positions} in \texttt{scf.out}. \\
    \hline
    \texttt{Dielectric Tensor} & Dielectric Tensor, which is a $3\times3$ matrix of real numbers. This parameter is optional, depending on whether LO-TO splitting needs to be considered in the system. \\
    \hline
    \texttt{Effective Charges} & Effective Charges, which are $3\times3$ matrices of real numbers for each of the atom. This parameter is optional, depending on whether LO-TO splitting needs to be considered in the system. \\
    \Xhline{4\arrayrulewidth}
  \end{tabularx}
  \label{table:dyn}
\end{table}

\begin{figure}[]
  \centering
  \includegraphics[width=1.0\linewidth]{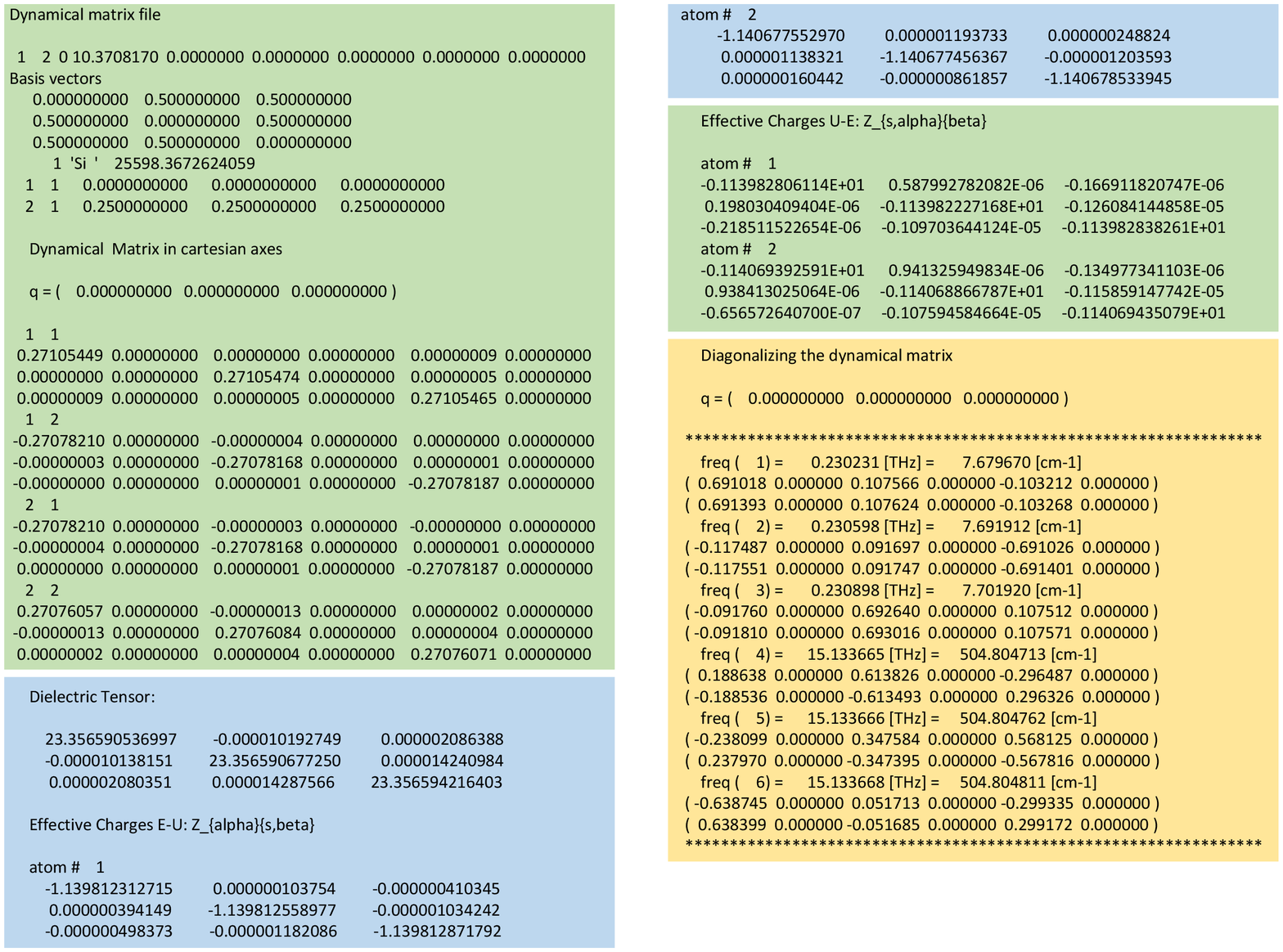}
  \caption{Harmonic phonon information at $\Gamma$ point of diamond silicon appears at the beginning of \texttt{dyn.out} file. The yellow part is required and the blue part is optional as \texttt{phq} program input, while the green part is not required and will not be read in by the program.}
  \label{fig:dyn}
\end{figure}

If other harmonic phonon calculation program is used, the \texttt{dyn.out} file should be prepared according to Table \ref{table:dyn}. It is further illustrated in Fig.~\ref{fig:dyn}, where the harmonic phonon information of the first {\bf q}-point ($\Gamma$ point) of diamond silicon is shown as an example. The yellow and blue parts are required and optional respectively as the program input while the green part is not required as input. Parameters that are required or optional need to be prepared by users, while parameters that are not required can be left blank. Note that the character strings \texttt{Dielectric}, \texttt{Effective}, \texttt{Diagonalizing}, \texttt{q} and \texttt{freq} are arguments that will be read in by the \texttt{phq} program and therefore should not be modified. \texttt{Dielectric Tensor} and \texttt{Effective Charges} are only needed to be provided in the harmonic phonon information of the first {\bf q}-point.

\subsection{md.out}
\begin{figure}[]
  \centering
  \includegraphics[width=1.0\linewidth]{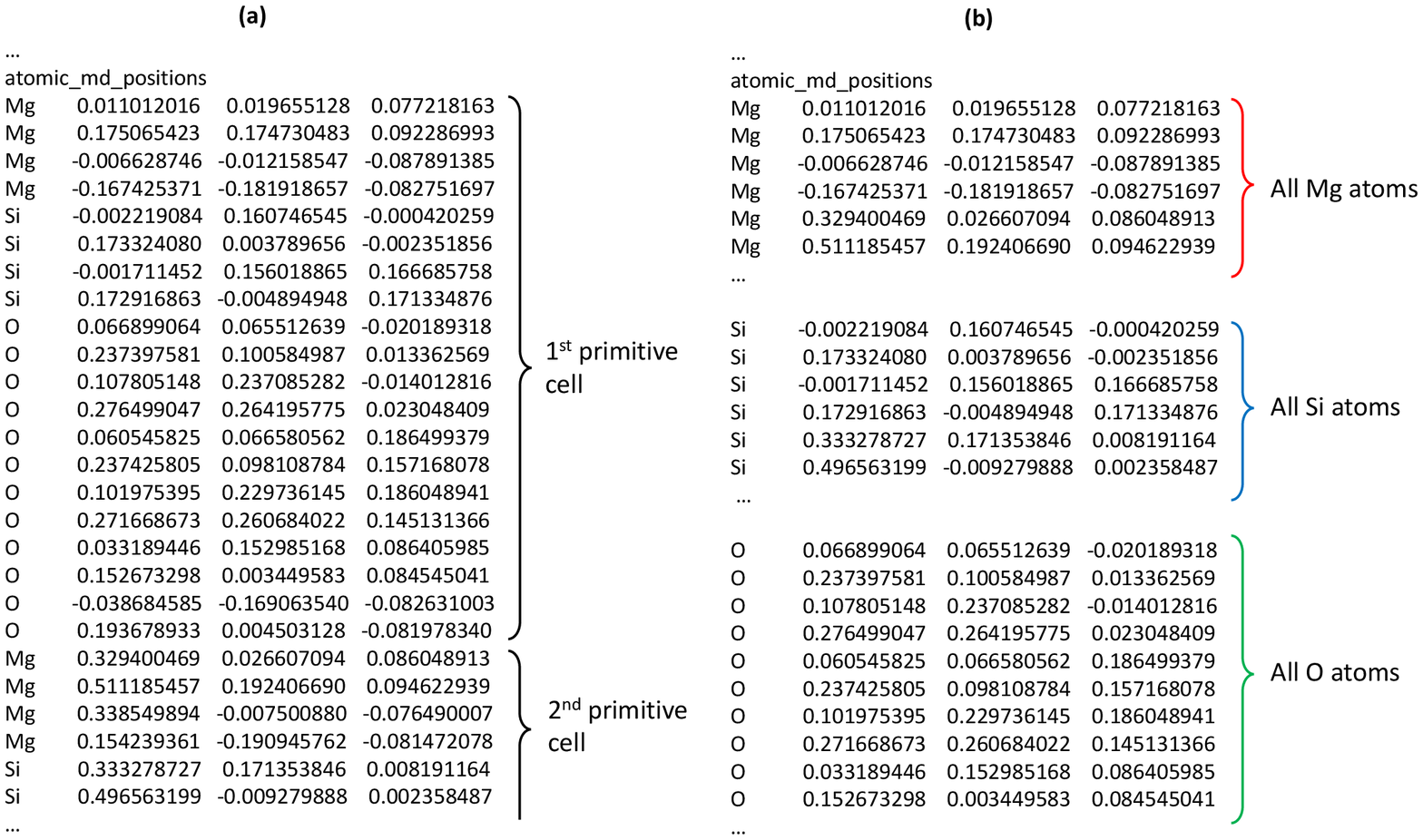}
  \caption{Atomic positions of $3\times3\times3$ supercell of \textit{Pbnm} MgSiO$_3$ are displayed in two formats: (a) Atoms that belong to the same primitive cell are put together. (b) Atoms of the same element are put together. \texttt{md.out} should take format (a) as \texttt{phq} program input.}
  \label{fig:md}
\end{figure}

\texttt{md.out} provides MD information. The initial atomic coordinates of the supercell before the MD run and the instantaneous atomic coordinates during the MD simulation need to be recorded. The details of the parameters are described in Table \ref{table:md}. Note that for either \texttt{atomic\_positions} or \texttt{atomic\_md\_positions}, atoms in each one primitive cell should be together instead of atoms of the same element being together. It is further illustrated in Fig.~\ref{fig:md}, which shows an example \texttt{md.out} file of $3\times3\times3$ supercell of \textit{Pbnm} MgSiO$_3$ in two formats. Fig.~\ref{fig:md}(a) takes the convention of Quantum ESPRESSO where atoms in one primitive cell are put together, while in Fig.~\ref{fig:md}(b), all Mg atoms, Si atoms and O atoms are put together. Here, format demonstrated in Fig.~\ref{fig:md}(a) should be taken for \texttt{md.out}. Also, the order of atoms in each of the primitive cell should be in the same order as provided in \texttt{scf.out}. The way to extend the primitive to generate the supercell for the MD simulation can be determined by the users.

\begin{table}[]
  \caption{Parameters used in the \texttt{md.out}}
  \centering
  \renewcommand{\arraystretch}{1.2}
  \begin{tabularx}{\textwidth}{l X}
    \Xhline{4\arrayrulewidth}
    Parameter name & Description \\
    \Xhline{3\arrayrulewidth}
    \texttt{total\_step} & Total actual MD steps. In practice, recorded MD steps should be configurations after reaching thermal equilibrium and therefore less than this number. \\
    \hline
    \texttt{atomic\_positions} & Initial atomic positions in reduced coordinates of the supercell lattice vectors. \\
    \hline
    \texttt{md\_step} & Integer numbers of recorded MD steps, followed by the instantaneous \texttt{atomic\_md\_positions} in the MD simulation. \\
    \hline
    \texttt{atomic\_md\_positions} & Atomic positions in reduced coordinates of the supercell lattice vectors during the MD simulation. \\
    \Xhline{4\arrayrulewidth}
  \end{tabularx}
  \label{table:md}
\end{table}

\section{Ouput files}
A detailed description of all the output files of \texttt{phq} program are summarized in Table \ref{table:output}. The main output files are \texttt{frequency.freq}, which records the renormalized frequencies, and \texttt{dynmatmd1}, ..., \texttt{dynmatmdnq}, which record the effective harmonic phonon information with the same format of \texttt{ph.x} executable's output and can be read in by the \texttt{q2r.x} executable in the Quantum ESPRESSO suite. Anharmonic force constants, renormalized phonon dispersions, anharmonic entropy and free energy can be obtained from \texttt{q2r.x} executable's output. If other Fourier interpolation program is used, either use the effective harmonic dynamical matrices file \texttt{dynamical\_matrix\_md.mat} or use the effective harmonic force constant matrix file \texttt{gamma\_matrix.mat} to fit the format.

\begin{table}[]
  \caption{Output files of the \texttt{phq} program}
  \centering
  \renewcommand{\arraystretch}{1.2}
  \begin{tabularx}{\textwidth}{l X}
    \Xhline{4\arrayrulewidth}
    Output & Description \\
    \Xhline{3\arrayrulewidth}
    \texttt{corr.vaf} & VAF of each eigenmode of the supercell. The VAF of each eigenmode lies on each column following the correlation time on the first column. \\
    \hline
    \texttt{corr\_fit.vaf} & Fitted exponentially decaying cosine functions (Eq.~(5)) of correlation functions according to Green's function from molecular dynamics \cite{Sun2010}. The fitted curve of VAF of each eigenmode of the supercell lies on each column following the correlation time on the first column. \\
    \hline
    \texttt{corr\_fit\_fourier.vaf} & FT of the fitted VAF curves. The FT of fitted VAF curve of each eigenmode of the supercell lies on each column following their common frequency range lies on the first column. \\
    \hline
    \texttt{corr\_fourier.vaf} & FT of VAFs. The FT of VAF of each eigenmode of the supercell following its frequency range lie on every two columns. \\
    \hline
    \texttt{dynamical\_matrix\_md.mat} & Effective harmonic dynamical matrices. It is an assemble of all the effective harmonic dynamical matrices from \texttt{dynmatmd1}, ..., \texttt{dynmatmdnq}. \\
    \hline
    \texttt{dynmatmd0} & {\bf q}-point list with the same format of \texttt{ph.x} executable's output in the Quantum ESPRESSO suite. They are in cartesian coordinates in unit of $2\pi$/\texttt{lattice\_parameter}. \\
    \hline
    \texttt{dynmatmd1}, ..., \texttt{dynmatmdnq} & Renormalized phonon information with the same format of \texttt{ph.x} executable's output in the Quantum ESPRESSO suite. nq is the total number of the {\bf q}-points. They share the same set of parameters of \texttt{dyn.out}. \\
    \hline
    \texttt{frequency.freq} & Harmonic, fitted renormalized, FT renormalized and MEM renormalized phonon frequencies lie on the second, third, fourth and fifth column separately while the first column labels different eigenmodes of the supercell. \\
    \hline
    \texttt{gamma\_matrix.mat} & Effective harmonic force constant matrix, which are $3\times3$ matrices of real numbers between each pair of atoms in the supercell. \\
    \hline
    \texttt{harmonic\_matrix.mat} & Harmonic force constant matrix, which are $3\times3$ matrices of real numbers between each pair of atoms in the supercell. \\
    \hline
    \texttt{msd.out} & Mean square displacement of each atom in the supercell. \\
    \hline
    \texttt{tau\_fit.tau} & Phonon quasiparticle's lifetime of each eigenmode of the supercell obtained by fitting the VAFs to Eq.~(5). \\
    \hline
    \texttt{tau\_fourier.tau} & Phonon quasiparticle's lifetime extracted from the FT of VAF of each eigenmode of the supercell. \\
    \hline
    \texttt{tau\_mem.tau} & Phonon quasiparticle's lifetime extracted from the MEM power spectrum of each eigenmode of the supercell. \\
    \hline
    \texttt{vector.out} & Eigenmodes of the supercell. \\
    \hline
    \texttt{vector\_q.out} & Eigenmodes of the primitive cell at each {\bf q}-point. \\
    \hline
    \Xhline{4\arrayrulewidth}
  \end{tabularx}
  \label{table:output}
\end{table}

\section{Example}
Here we illustrate the performance of the \texttt{phq} code using diamond Si as an example. Anharmonic phonon dispersion, vibrational density of states, and anharmonic free energy are obtained. We use the density-functional theorem based Vienna \textit{ab initio} simulation package (VASP)~\cite{vasp1,vasp2} to carry out MD simulations on $4\times4\times4$ supercells (128 atoms) of Si. We employ the generalized gradient approximation of Perdew, Burke, and Ernzerhof~\cite{pbe} and the projector-augmented wave method~\cite{paw} with an associated plane-wave basis set energy cutoff of 320 eV. Simulations are carried out at the volume with a zero static pressure.  Temperatures ranging from 300 K to 1500 K are controlled through the Nos\'e dynamics~\cite{nose}. The MD runs for 50 ps with a time step of 1 fs.  Harmonic phonons are calculated using density-functional perturbation theory (DFPT)~\cite{dfpt}.

\subsection{Obtaining VAFs}

\begin{figure}[]
  \centering
  \includegraphics[width=1.0\linewidth]{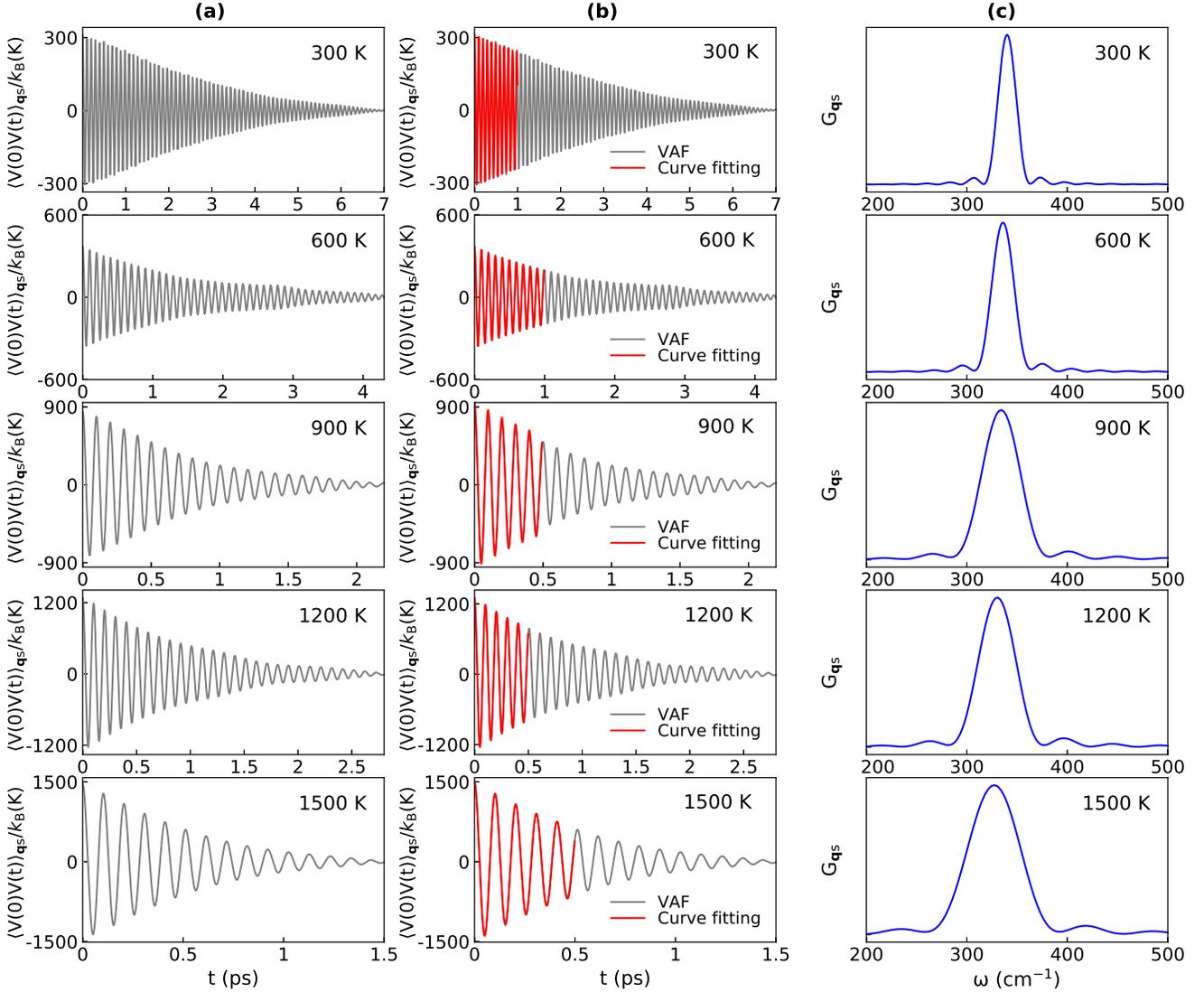}
  \caption{Mode-projected velocity autocorrelation functions (VAF) (grey) of the LA/LO mode at ${\bf q}=W$ of diamond silicon at 300, 600, 900, 1200 and 1500 K respectively are shown in (a) and (b). The corresponding power spectra (blue) are shown in (c). (b) shows in red the curves fitted with Eq.~(5) for comparison.}
  \label{fig:vaf}
\end{figure}

The mode-projected VAFs, $\langle V_{{\bf q}s}(0) V_{{\bf q}s}(t) \rangle$, are given by \texttt{corr.vaf} file. Fig. \ref{fig:vaf}(a) displays the calculated $\langle V_{{\bf q}s}(0) V_{{\bf q}s}(t) \rangle$ at several temperatures of the LA/LO mode at ${\bf q}=W$ ([1/4, 3/4, 1/2]). At each temperature, the VAF amplitude decays exponentially. Accordingly, its power spectrum obtained via a Fourier transformation (FT) of VAFs has a Lorentzian line shape with a main single peak, Fig. \ref{fig:vaf}(c).  This behavior demonstrates the validity of the concept of phonon quasiparticle and allows us to identify the renormalized phonon frequency ($\widetilde{\omega}_{{\bf q}s}$) and line width ($\widetilde{\Gamma}_{{\bf q}s}$) of the phonon mode. Note that the file recording the FT is \texttt{corr\_fourier.vaf}.

In practice, however, it is not a optimal way to find $\widetilde{\omega}_{{\bf q}s}$ and $\widetilde{\Gamma}_{{\bf q}s}$ directly from the power spectrum obtained from FT. Instead, we fit the VAFs to an exponentially decaying cosine function \cite{Sun2010, pgm2, Lu2017, HofScience} to extract $\widetilde{\omega}_{{\bf q}s}$ and  $\widetilde{\Gamma}_{{\bf q}s}$, Eq.~(5). From our previous work~\cite{Zhang2014, Zhang2017, Lu2017}, $\widetilde{\omega}_{{\bf q}s}$ and  $\widetilde{\Gamma}_{{\bf q}s}$ can be obtained by performing the fitting only for the first few oscillation periods of VAFs. Therefore, to do the fitting, one only needs to calculate VAFs for a relatively short correlation time. Note that the file recording the fitting is \texttt{corr\_fit.vaf}.

The fitted curve at each temperature is shown in Fig. \ref{fig:vaf}(b), matching the VAF very well. Our code also offers a way to obtain $\widetilde{\omega}_{{\bf q}s}$ by maximum entropy method (MEM)~\cite{dynaphopy}.  A list of $\widetilde{\omega}_{{\bf q}s}$ obtained by different methods can be found in the \texttt{frequency.freq} file. Users can choose renormalized frequencies obtained by fitting approach, FT or MEM to obtain the effective force constants and dynamical matrices by specifying 0, 1 or 2 for the \texttt{method} in the \texttt{input} file.

\subsection{Obtaining frequency shifts}

\begin{figure}[h]
  \centering
  \includegraphics[width=1.0\linewidth]{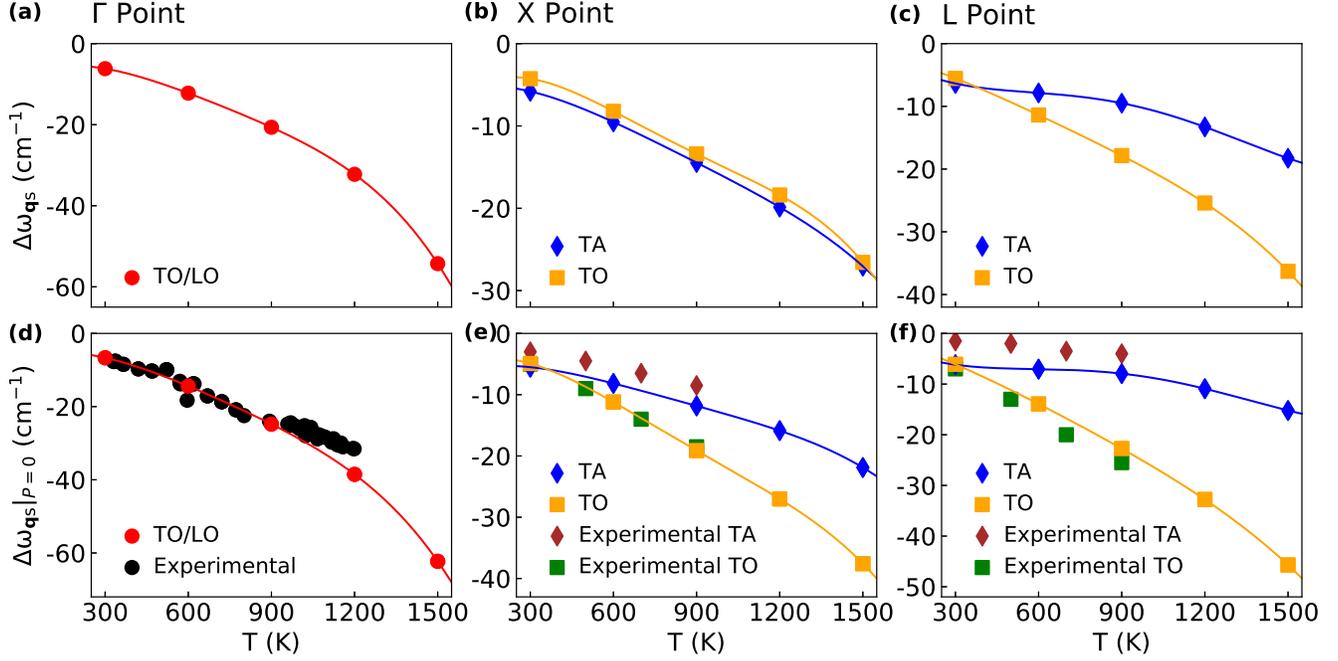}
  \caption{Temperature dependent frequency shifts $\Delta\omega_{{\bf q}s}$ of diamond Si with a volume where the static pressure is zero of: (a) TO/LO mode at $\Gamma$,  (b) TA and TO mode at $X$, and (c) TA and TO mode at $L$ high-symmetry point. (d)-(f) show the corresponding corrected  frequency shifts at zero pressure, $\Delta\omega_{{\bf q}s}|_{P=0}$, at different temperatures. Experimental results obtained by Raman spectra \cite{si1, si2} are shown for comparison.}
  \label{fig:omega}
\end{figure}

As a demonstration of the validity of the present approach, we compare the calculated thermal frequency shifts given by \texttt{frequency.freq} file with experimental measurements. Fig. \ref{fig:omega}(a)-\ref{fig:omega}(c) display frequency shifts $\Delta\omega_{{\bf q}s} = \widetilde{\omega}_{{\bf q}s}|_{V=V_{eq}} - \omega_{{\bf q}s}$ of a TO/LO mode at the $\Gamma$ point, TA and TO modes at the $X$ point ([0, 1/2, 1/2]), and at the $L$ point ([1/2, 1/2, 1/2]) at the equilibrium volume of zero static pressure. Because experiments were conducted under ambient pressure, the renormalized phonon frequencies obtained under constant volume conditions have to be corrected. We indicate that such correction can be properly captured by considering the thermal shifts caused by the temperature induced volume expansion \cite{Zhang2014},
\begin{equation}
\widetilde{\omega}_{{\bf q}s}(T)|_{P=0} = \widetilde{\omega}_{{\bf q}s}(T)|_{V=V_{eq}} + \omega_{{\bf q}s}[{\rm exp}(-
\gamma_{{\bf q}s}\bar{\alpha}T)-1],
\end{equation}
where $\gamma_{{\bf q}s}$ is the mode Gr\"{u}neisen parameter and $\bar{\alpha}$ is the temperature averaged thermal expansion coefficient:
\begin{equation}
\bar{\alpha}(T) = \frac{1}{T} \int_{0}^{T}\alpha(T')dT'.
\end{equation}
The mode Gr\"{u}neisen parameter connects the phonon frequency variation with the volume expansion and can be obtained by DFPT \cite{dfpt}. It is defined as \cite{gruneisen1}:
\begin{equation}
\gamma_{{\bf q}s} = -\frac{d{\rm ln}\omega_{{\bf q}s}(T)}{d{\rm ln}V(T)}.
\end{equation}
In practice the weak temperature dependence can be ignored \cite{gruneisen2}:
\begin{equation}
\gamma_{{\bf q}s} = -\frac{d{\rm ln}\omega_{{\bf q}s}(0)}{d{\rm ln}V(0)},
\end{equation}
where $\omega_{{\bf q}s}(0)$ is the harmonic phonon frequency at volume $V(0)$. $\alpha(T)$ is the thermal expansion coefficient which can be obtained by QHA. It is defined as:
\begin{equation}
\alpha(T) = \frac{1}{V(T)}\frac{dV}{dT}.
\end{equation}
With this, according to Eq.~(11), the temperature averaged thermal expansion coefficient can be onbtained:
\begin{equation}
\bar{\alpha}(T) = \frac{1}{T}{\rm ln}\frac{V(T)}{V(0)}.
\end{equation}
Overall, our calculated $\Delta\omega_{{\bf q}s}|_{P=0} = \widetilde{\omega}_{{\bf q}s}|_{P=0} - \omega_{{\bf q}s}$ are in good agreement with experiments \cite{si1, si2}, Fig. \ref{fig:omega}(d)-\ref{fig:omega}(f). We note that at the $X$ point, TA and TO modes have similar frequency shifts at constant volume. After the incorporation of the volume expansion, i.e., at constant pressure, their frequency shifts differ significantly. This is essentially caused by the different Gr\"{u}neisen parameters of these two modes.

\subsection{Obtaining temperature dependent phonon dispersion}

Temperature dependent renormalized phonon frequencies obtained as such are only for those few {\bf q}-points sampled by MD simulations, which are not sufficient to converge the calculation of thermal properties. Nevertheless, these frequencies provide a basis to obtaining phonon frequencies at {\bf q}-points throughout the Brillouin zone. Here, we rely on an effective harmonic dynamic matrix, Eq.~(6), and the derived effective harmonic force constants, Eq.~(7), which give us the effective harmonic dynamic matrix at arbitrary {\bf q}-points via a Fourier interpolation, Eq.~(8) \cite{Zhang2014}.

To construct the  effective harmonic dynamic matrix along this line, users first find files of \texttt{dynmatmd0}, \texttt{dynmatmd1}, ..., \texttt{dynmatmd64} given by \texttt{phq}, and copy them to the \texttt{phq/example/Si/postprocessing/dispersion} sub-folder. Files \texttt{q2r.in}, \texttt{dispersion.in} and \texttt{plotband.in} files are also there. Next run:
\\
\texttt{q2r.x < q2r.in > q2r.out}
\\
\texttt{matdyn.x < dispersion.in > dispersion.out}
\\
\texttt{plotband.x < plotband.in > plotband.out}
\\
to calculate anharmonic phonon dispersion of diamond silicon, where \texttt{matdyn.x} and \texttt{plotband.x} are also Quantum ESPRESSO executables.

\begin{figure}[h]
  \centering
  \includegraphics[width=0.6\linewidth]{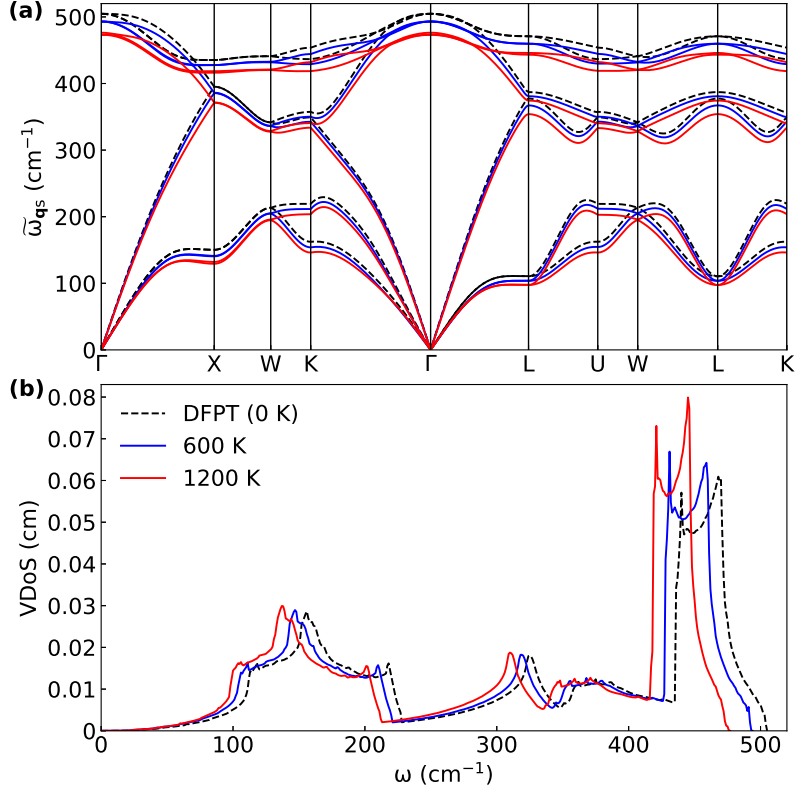}
  \caption{(a) Anharmonic phonon dispersions at 600 K (blue solid curve) and 1200 K (red solid curve) compared with harmonic phonon dispersion calculated by density functional perturbation theory (DFPT) (black dashed curve) of diamond Si at the volume with a zero static pressure. (b) Temperature dependent vibrational density of states (VDoS) of diamond Si obtained on $20\times20\times20$ {\bf q}-point grid at 600 K (blue solid curve), 1200 K (red solid curve). The VDoS at 0 K obtained with DFPT (black dashed curve) is shown for comparison.}
  \label{fig:dispersion}
\end{figure}

Fig. \ref{fig:dispersion} (a) displays the obtained anharmonic phonon dispersions at 600 K and 1200 K. The harmonic phonon dispersions calculated by DFPT at static zero-pressure equilibrium volume is also shown for comparison. It can be seen that the temperature effect on phonon frequencies is discernible, and the overall trend is that phonon frequencies decrease as temperature increases~\cite{fultz}. Accordingly, the major peaks of the obtained vibrational density of states (VDoS) shift to lower frequencies with increasing temperature, Fig. \ref{fig:dispersion} (b). Note that the VDoSs are calculated with a $20\times20\times20$ {\bf q}-point grid, equivalent to a supercell of 16000 atoms in a direct MD simulation, which is well beyond the capability of \textit{ab initio} methods.

\subsection{Obtaining anharmonic vibrational entropy and free energy}

With the renormalized phonon frequencies, the entropy is obtained by an integration over the whole Brillouin zone \cite{qha2},
\begin{equation}
S = k_B \sum_{{\bf q}s} [(n_{{\bf q}s} + 1){\rm ln}(n_{{\bf q}s} + 1) - n_{{\bf q}s}{\rm ln}n_{{\bf q}s}] ,
\end{equation}
where $n_{{\bf q}s}=[{\rm exp}(\hbar\widetilde{\omega}_{{\bf q}s}/k_BT)-1]^{-1}$.
The free energy is obtained by numerical integration \cite{Zhang2014}:
\begin{equation}
F(T) = E_0 + \frac{1}{2} \sum_{{\bf q}s} \hbar \omega_{{\bf q}s} - \int_{0}^{T} S(T')dT',
\end{equation}
where $E_0$ is the static ground state energy and $\omega_{{\bf q}s}$ is the harmonic phonon frequency.

It can be seen that $\omega_{{\bf q}s}$ is key quantity to calculate $S$ and $F$. To obtain $\omega_{{\bf q}s}$ over the whole Brillouin zone, users copy the \texttt{dynmatmd0}, \texttt{dynmatmd1}, ..., \texttt{dynmatmd64} files obtained from \texttt{phq} to the \texttt{phq/example/Si/postprocessing/vdos} sub-folder. \texttt{q2r.in} and \texttt{vdos.in} files are also there. Then, run:
\\
\texttt{q2r.x < q2r.in > q2r.out}
\\
\texttt{matdyn.x < vdos.in > vdos.out}
\\
to calculate anharmonic VDoS of diamond silicon on a $20\times20\times20$ {\bf q}-point grid. $S$ and $F$ can be readily obtained from output VDoS file \texttt{vdos} according to Eq.~(16)(17).

\begin{figure}[]
  \centering
  \includegraphics[width=0.6\linewidth]{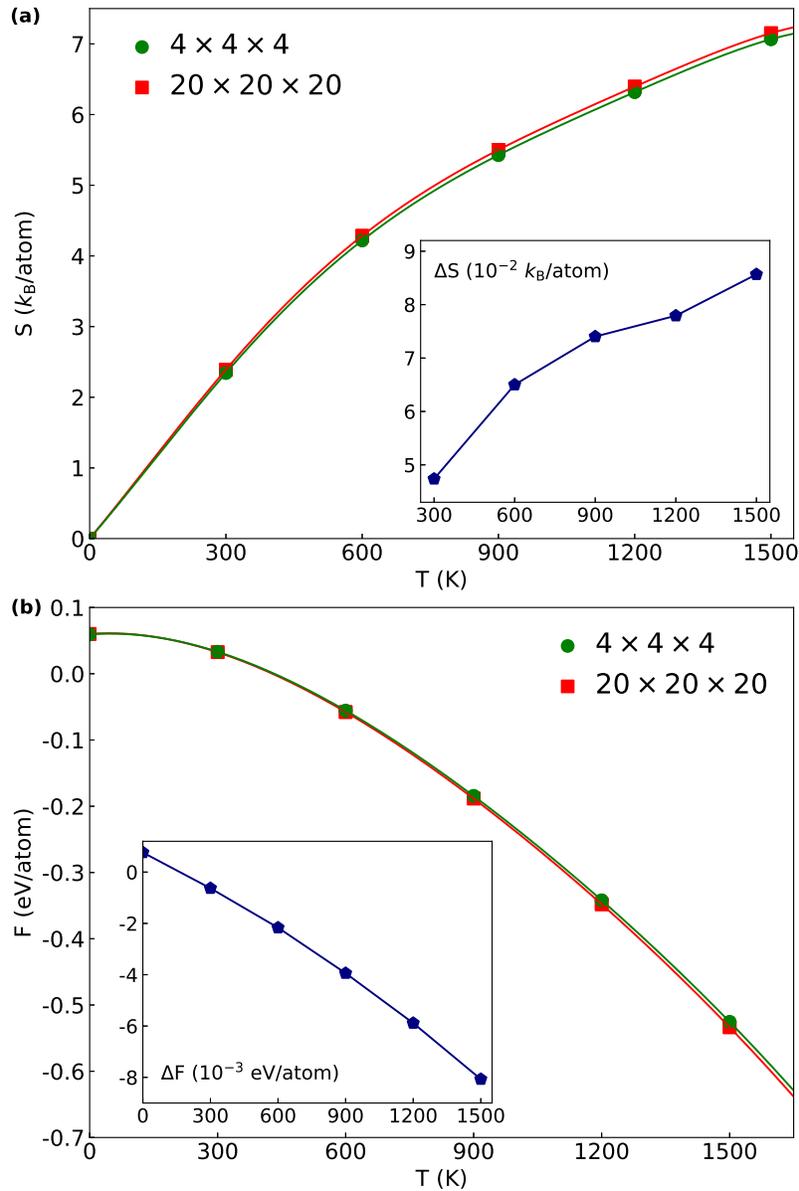}
  \caption{ (a) Vibrational entropy ($S$) and (b) vibrational free energy ($F$) of diamond Si at volume with zero static pressure calculated with different {\bf q}-grids: $4\times4\times4$ and $20\times20\times20$. Inserts: (a) Differences in vibrational entropy ($\Delta S$) and (b) vibrational free energy ($\Delta F$) obtained with $4\times4\times4$ and $20\times20\times20$ {\bf q}-grids.}
  \label{fig:entropy}
\end{figure}

Fig. \ref{fig:entropy} compares $S$ and $F$ obtained with different {\bf q}-grids: $4\times4\times4$ and $20\times20\times20$.
The discernible differences $\Delta S$, Fig. \ref{fig:entropy}(a), and thus $\Delta F$, Fig. \ref{fig:entropy}(b) reveal that it is necessary to use a sufficiently large number of {\bf q}-point in order to converge thermal properties.

\section{Conclusion}

In this communication we report \texttt{phq}, a code for extracting phonon quasiparticle properties, obtaining temperature dependent renormalized phonon dispersions, and calculating anharmonic free energies from harmonic phonon calculations and MD simulations. The program can be installed on Linux, macOS and Windows operating systems. Outputs from VASP \cite{vasp1, vasp2} and Quantum ESPRESSO \cite{qe} are both accepted as inputs here.  Extension of compatibility with ABINIT \cite{abinit}, LAMMPS \cite{lammps}, Phon \cite{phon} and Phonon \cite{phonon} for phonon dispersions processing are planned for the near future.

\section*{Acknowledgments}

This work was supported primarily by NSF grant EAR-1503084. This work used the Extreme Science and Engineering Discovery Environment (XSEDE), which is supported by National Science Foundation grant number TG-DMR170058.
DBZ acknowledges support from NSFC grant 11874088 and U1530401. TS is supported by NSFC 41474069. Computations are performed at the Stampede2 (the flagship supercomputer at the Texas Advanced Computing Center (TACC), University of Texas at Austin) and at Beijing Computational Science Research Center.

\begin{appendices}

\section{Fourier transformation}
\label{appendix:ft}
The Fourier transformation (FT) of function $c(t)$ ($t\geq0$),
\begin{equation}
C(\omega) = \int_{0}^{\infty} c(t) e^{i \omega t} dt
\end{equation}
is done by mapping the $N$ consecutive evenly sampled points in the time domain to the frequency domain
\begin{equation}
D_k = \sum_{j=0}^{N-1}c_j \omega_j e^{i2\pi jk/N} \ \ \ \ \ \ k=0, ..., N-1,
\end{equation}
where $\omega_j$ is the window function. Then the discrete power spectrum is defined at $N$ nonnegative frequencies as
\begin{equation}
P(f_k) = \frac{1}{N \sum_{j=0}^{N-1}\omega_j^2} \left| D_k \right|^2.
\end{equation}
$f_k$ is defined only for the zero and positive frequencies
\begin{equation}
f_k = \frac{k}{N\Delta} = f_c \frac{k}{N} \ \ \ \ \ \ k=0, ..., N-1,
\end{equation}
where $f_c$ is the upper limit of the frequency, known as Nyquist frequency. The actual frequency range considered in the calculation is limited by the window function. In \texttt{phq} program, by default the upper limit of the window function of each mode is set to be integral period where the amplitude of the velocity autocorrelation function (VAF) drop to 0.75 of its maximum. This value can be modified by users in subroutine \texttt{correlation} of the \texttt{main.f90} file. By default, rectangular window function is used
\begin{equation}
\omega_j = 1 \ \ \ \ \ \ j=0, ..., N-1,
\end{equation}
which is ready to obtain power spectrum of Lorentzian line shape. To avoid frequency leakage into neighboring bins, Hann window is also provided
\begin{equation}
\omega_j = {\rm sin}\left[\frac{\pi (N-j)}{2N}\right]^2 \ \ \ \ \ \ j=0, ..., N-1.
\end{equation}
Users can enable Hann window in subroutine \texttt{corr\_fourier}.

\section{Maximum entropy method}
\label{appendix:mem}
Maximum entropy method (MEM) is also known as all-poles model, autoregressive model, which is used to obtain smooth power spectrum in signal processing. Here we assume that the power spectrum of each mode follows the approximation
\begin{equation}
P(f) = \frac{a_0}{\left| 1+\sum_{j=1}^{N} a_j e^{i 2\pi f \Delta j} \right|^2},
\end{equation}
where $a_0$ as well as $a_j$ is a set of coefficients of this approximation and $\Delta$ is the sampling interval in the time domain. $a_0$ and $a_j$ can be obtained from linear prediction method (LP)~\cite{recipes}
\begin{equation}
y_n = \sum_{j=1}^{M} a_j y_{n-j} + x_n
\end{equation}
where $y_n$ are a series of input and also predicted values by the LP, $x_n$ is the discrepancy between the data and the predicted value. Here we use the polarized velocities as the input of LP. A recursive procedure called Burg method~\cite{recipes} is adopted to minimize the discrepancy $x_n$, which is performed in subroutine \texttt{linear\_response} of the \texttt{main.f90} file. Regarding it to take input $x_n$ into output $y_n$, LP has a filter function
\begin{equation}
\mathcal{H}(f) = \frac{1}{1-\sum_{j=1}^{M}d_j e^{i 2\pi f \Delta j}}
\end{equation}
Thus, the power spectrum of the $y_n$ equals the power spectrum of the $x_n$ multiplied by $\left| \mathcal{H}(f) \right|^2$. Therefore, the coefficients $a_0$ and $a_j$ can be related to the LP coefficients by
\begin{equation}
a_0 = \texttt{xms}, \ \ \ a_j = -\texttt{d}(j) \ \ \ \ \ \ j=1, ..., N.
\end{equation}
Then the MEM power spectrum can be obtained from Eq.~(24). Phonon quasiparticle properties are extracted from the power spectrum, which is done in subroutine \texttt{maximum\_entropy}. A fitting of the MEM power spectrum of each mode to a Lorentzian line shape performed in subroutine \texttt{lorentzian} also give quasiparticle properties for comparison.

\end{appendices}








\end{document}